\documentclass[english,aps, twocolumn]{revtex4}
\usepackage[T1]{fontenc}
\usepackage[latin9]{inputenc}
\usepackage{amssymb}
\usepackage{graphicx}
\usepackage{esint}

\makeatletter
\@ifundefined{textcolor}{}
{%
 \definecolor{BLACK}{gray}{0}
 \definecolor{WHITE}{gray}{1}
 \definecolor{RED}{rgb}{1,0,0}
 \definecolor{GREEN}{rgb}{0,1,0}
 \definecolor{BLUE}{rgb}{0,0,1}
 \definecolor{CYAN}{cmyk}{1,0,0,0}
 \definecolor{MAGENTA}{cmyk}{0,1,0,0}
 \definecolor{YELLOW}{cmyk}{0,0,1,0}
}

\makeatother

\usepackage{babel}
\begin{document}

\title{Transfer of Vibrational Coherence Through Incoherent Energy Transfer
Process in F\"{o}rster Limit}

\author{Tom\'{a}\v{s} Man\v{c}al$^{1}$, Jakub Dost\'{a}l$^{1,2}$, Jakub
P\v{s}en\v{c}\'{i}k$^{1}$, and Donatas Zigmantas$^{2}$}

\affiliation{$^{1}$Faculty of Mathematics and Physics, Charles University in
Prague, Ke Karlovu 5, CZ-121 16 Prague 2, Czech Republic}

\affiliation{$^{2}$Department of Chemical Physics, Lund University, Getingev\"{a}gen
60, 221 00 Lund, Sweden}
\begin{abstract}
We study transfer of coherent nuclear oscillations between an excitation
energy donor and an acceptor in a simple dimeric electronic system
coupled to an unstructured thermodynamic bath and some pronounced
vibrational intramolecular mode. Our focus is on the non-linear optical
response of such a system, i.e. we study both excited state energy
transfer and the compensation of the so-called ground state bleach
signal. The response function formalism enables us to investigate
a heterodimer with monomers coupled strongly to the bath and by a
weak resonance coupling to each other (F\"{o}rster rate limit). Our
work is motivated by recent observation of various vibrational signatures
in 2D coherent spectra of energy transferring systems including large
structures with a fast energy diffusion. We find that the vibrational
coherence can be transferred from donor to acceptor molecules provided
the transfer rate is sufficiently fast. The ground state bleach signal
of the acceptor molecules does not show any oscillatory signatures,
and oscillations in ground state bleaching signal of the donor prevail
with the amplitude which is not decreasing with the relaxation rate. 

\end{abstract}
\maketitle

\section{Introduction}

Ultrafast time-resolved non-linear spectroscopy of electronic transitions
represents an indispensable tool for the study of photoinduced dynamic
and kinetic processes in wide range of interesting molecular and solid
state systems \cite{MIlota2009a,Cheng2009a,Ginsberg2009a,Collini2009a,Turner2010a,Collini2010a,Schlau-Cohen2011a,Wong2012a,Turner2012c,Bixner2012a,Dostal2012a}.
While the degrees of freedom (DOF) of the studied systems that are
directly addressed in these experiments are electronic, fine details
of the time resolved spectra depend crucially on the characteristics
of the surrounding nuclear modes. It was recognized early on that
some ultrafast techniques yield unprecedented time-dependent information
about the nuclear modes, eventhough these might form a thermodynamic
bath with broad spectral density. For instance, the so-called photon
echo peakshift experiment yields almost directly the form of the energy
gap correlation (also termed bath correlation function) \cite{Cho1996a,Yang1999b,Mancal2004a}.
With the advent of the two-dimensional (2D) coherent Fourier transformed
spectroscopy \cite{Brixner2004a,Brixner2005a}, it was hoped shortly
that one obtained a method which would provide insight into electronic
coupling between chromophores forming molecular complex. Instead,
the strength of the method proved to be in allowing observation of
coherent oscillatory features of electronic and nuclear origin \cite{Kjellberg2005a,Pisliakov2006a,Engel2007a,Egorova2008a,Nemeth2008a,Collini2009a,Collini2010a,Turner2011a,Caram2012a}.
In particular, large effort has been made to understand the vibrational
features revealed by the 2D spectroscopy and to contrast them to the
electronic features \cite{Turner2011a,Mancal2011b,Butkus2012a,Butkus2012b,seibt2013a}.
Motivation for this work can be found in the ongoing debate on the
origin of the long lived coherent oscillations observed in the 2D
spectra of some molecular complexes \cite{Engel2007a,Collini2009a,Panitchayangkoon2010a,Collini2010a}
for which models employing nuclear vibrational modes are also suggested
\cite{Christensson2012a,Tiwari2013a,Chin2013a,Chenu2013a}. Setting
aside this particular debate, the study of the vibrational features
of 2D spectra has its own importance. The vibrational DOF are ubiquitous
in molecular aggregates and almost all ultrafast time-resolved spectroscopic
methods are sensitive to them. Notable exception would be an ideal
frequency integrated pump-probe spectroscopy for which one can show
that it is insensitive to the nuclear vibrations \cite{Yuen-Zhou2012a}.
2D spectroscopy, however, can be shown to have a specific sensitivity
to vibrational features \cite{Mancal2012b}, and the frequency resolved
pump-probe spectroscopy with finite pulses keeps sufficient sensitivity
to vibrational features, too. 

Our theoretical work in this paper is motivated by our recent observation
of coherent oscillations in low temperature 2D spectra of chlorosome
\cite{dostal2013a}. Previous measurements by pump-probe technique
on this system showed convincingly that nuclear oscillations occur
in the electronic ground state after an ultrafast excitation \cite{Ma1999a}.
The chlorosome is the largest known bacterial photosynthetic antenna,
containing $\sim10^{5}$ aggregated chlorophylls which are subject
to large disorder in optical gaps. Correspondingly, a plausible explanation
of the oscillations observed in 2D spectra at low temperature are
nuclear oscillations, as these could survive the disorder in the electronic
transition energies. In our previous room temperature measurements
of the chlorosome \cite{Dostal2012a}, we observed an initial ultrafast
time scale which could be associated with the fast diffusion of the
excitation between certain coherent domains formed in the disordered
energetic landscape of the chlorosome. We are therefore motivated
to study the influence of an ultrafast energy transfer on the evolution
of the nuclear oscillatory features in the 2D spectra. 

In order to understand the properties of nuclear oscillations observed
by 2D spectroscopy of photosynthetic energy transferring systems,
one needs to know spectroscopic properties and time-dependent signatures
of nuclear vibrations of at least the two limiting cases in which
these systems occur, namely the case of the F\"{o}rster energy transfer
between spatially localized excitons and the case of the delocalized
Frenkel excitons. In photosynthetic energy transferring antennae,
these cases never occur in their pure form as many photosynthetic
antennae fall in between the two limiting regimes. Nevertheless, when
trying to understand experimentally observed behaviors, it is useful
to understand spectroscopic signatures which could be assigned these
two limiting cases. In this paper, we start with the simpler of the
two regimes, namely with the F\"{o}rster energy transfer between
spatially localized excitations. In regard to our previous work, this
corresponds to the energy transfer between the coherent domains of
the chlorosome \cite{Dostal2012a}.

To simplify the theoretical treatment, we study a weakly coupled dimeric
system which possesses one pronounced intramolecular nuclear vibrational
mode with sufficient Huang-Rhys factor to be observable by electronic
non-linear spectroscopy, coherent 2D electronic spectroscopy in our
case. We avoid the situation of a resonance between the vibrational
frequency and the donor-acceptor energy gap in which the selected
intramolecular nuclear vibrational mode would influence the energy
transfer rate. This dimer interacts with its environment (solvent,
proteins surroundings etc.). We ask the question whether the vibrational
coherence excited by a sequence of two short pulses (the first two
of the three pulses of the 2D electronic spectroscopy technique) on
one molecule (denoted as donor here) can be transferred to a neighboring
molecule (an acceptor). We also study the fate of the oscillations
induced on the donor after it relaxes to the electronic ground state
when its excitation is transferred to the acceptor. The answers to
these questions could in principle shed some light on the behavior
of the coherent oscillations observed in chlorosomes.

The paper is organized as follows. In the next section we introduce
the model molecular Hamiltonian and we discuss the model of system-bath
interaction. We introduce the weak inter molecular coupling limit
- the F\"{o}rster limit. In Section \ref{sec:Non-linear-Response-Functions}
we derive response functions for all the signal components of the
coherent 2D spectra of a dimer system with a weak resonance coupling,
and we discuss their properties. In Section \ref{sec:Discussion}
we discuss particular numerical results and the dependence of the
amplitude of the transferred vibrational coherence on the energy transfer
rates. We present our conclusions in Section \ref{sec:Conclusions}.

\section{Molecular Model\label{sec:Molecular-Model}}

In this work we consider a molecular dimer. We will describe unidirectional
energy transfer from one molecule to another. We therefore denote
the molecule which we consider to be the excitation donor by a letter
$D$, while the other molecule (the excitation acceptor) will be denoted
by $A$. In actual molecular systems, both molecules can be donors
and acceptors, e.g. when they have similar transition frequency and
the energy transfers in one or the other way are close to equally
probable. A general theory of excitation energy transfer has to account
for the possible back transfer of an excitation back to the donor
molecule. However, as will be shown later, the oscillation amplitude
decays during the energy transfer, and it is enough to demonstrate
this on a single step of the energy transfer. In systems in which
the transition frequency of the acceptor molecule will be similar
to that one of the donor, such as in chlorosome, the probability of
the back transfer will be diminished by the presence of other possible
neighbors. Most of the relevant conclusions about the likelihood of
vibrational coherence transfer can therefore be reached from studying
unidirectional energy transfer. The theory developed below applies
directly to a hetero dimeric system where back transfer is small due
to large energy difference between the donor and acceptor.

We will describe the donor (acceptor) molecule in the dimer as a two-level
system with electronic ground states $|g_{D}\rangle$ ($|g_{A}\rangle$)
and excited state $|e_{D}\rangle$ ($|e_{A}\rangle$). To describe
a complete non-linear spectrum of such a system, it might be necessary
to add even higher lying excited states or band of states $|f_{D}\rangle$
($|f_{A}\rangle$). The two molecules interact through a resonance
coupling $J$ so that the electronic excitation can be exchanged between
them. This means that the electronic states of the individual molecules
are not eigenstates of the dimer. When, however, the interaction with
the environment is stronger than the resonance coupling, the identity
of the molecules is approximately preserved. Any delocalization possibly
established by resonance coupling is destroyed by fluctuations induced
by the environment. This is the limit of a strong bath influence,
$\lambda_{{\rm bath}}>J$, where $\lambda_{{\rm bath}}$ is the bath
reorganization energy - the parameter characterizing the system-bath
coupling. In the strong system-bath coupling limit, the appropriate
description of the energy transfer process is the F\"{o}rster rate
theory and its modifications \cite{Foerster1948a,Sumi1999a,Jang2004a}.
The F\"{o}rster rate can be calculated from the known absorption
and fluorescence spectra of the donor and acceptor molecules, respectively,
and it carries a prefactor proportional to $|J|^{2}$. In this work
we will be interested in the influence the value of the rate has on
the transfer of the vibrational coherence. We will not calculate the
rates, rather, we will use them as a free parameter assuming that
different rates correspond to different values of $J$. The dynamics
of the bath, which determines the line shape of the absorption and
the fluorescence, will be specified in terms of the bath correlation
function, also known as energy gap correlation function. 

The total Hamiltonian of the dimer reads as
\begin{equation}
H=H_{{\rm bath}}+\sum_{i=A,D}H_{i}^{(1)}+J(|A\rangle\langle D|+|D\rangle\langle A|)+H^{(2)},
\end{equation}
where $H_{{\rm bath}}$ is the Hamiltonian of the bath, $H_{i}^{(1)}$
are the Hamiltonians corresponding to the singly excited states of
non-interacting dimer $H_{i}=(\epsilon_{i}+\Delta V_{i})|i\rangle\langle i|,\; i=A,D$,
the states $|A\rangle$ and $|D\rangle$ are the singly excited states
of the non-interacting dimer
\begin{equation}
|A\rangle=|e_{A}\rangle|g_{D}\rangle,
\end{equation}
\begin{equation}
|D\rangle=|g_{A}\rangle|e_{D}\rangle,
\end{equation}
and $H^{(2)}=(\epsilon_{A}+\epsilon_{B}+\Delta V_{A}+\Delta V_{D})|e_{A}\rangle|e_{D}\rangle\langle e_{D}|\langle e_{A}|$
is the Hamiltonian of the doubly excited state of the dimer. If needed
it may include the higher excited states $|f_{D}\rangle$ and $|f_{A}\rangle$
of the donor and acceptor, respectively. The energy gap operator $\Delta V_{A}$
and $\Delta V_{B}$ describe the interaction of the acceptor and donor
with the their surrounding environment, respectively. We set the ground
state electronic energy $\epsilon_{g}$ to zero. We will ignore the
doubly excited state in our treatment, because for weak coupling $J$
(which is assumed in our F\"{o}rster type treatment) the excited
state absorption cancels with the ground state contributions which
would otherwise lead to crosspeaks in 2D spectra \cite{Mancal2013z}. 

The description of the electron-phonon coupling (the coupling of the
electronic states $|A\rangle$ and $|D\rangle$ with the bath of nuclear
degrees of freedom) will be done in terms of the bath correlation
functions $C_{A}(t)={\rm tr}\{\Delta V_{A}(t)\Delta V_{A}(0)W_{eq}^{(A)}\}$
and $C_{D}(t)={\rm tr}\{\Delta V_{D}(t)\Delta V_{D}(0)W_{eq}^{(D)}\}$
which describe the fluctuation of the transition energy on the acceptor
and the donor, respectively. The time argument on the energy gap operators
in the definitions of the bath correlation functions denotes interaction
picture with respect to the bath Hamiltonian and $W_{eq}^{(A)}$ ($W_{eq}^{(D)})$
denotes the equilibrium density operator of the acceptor (donor).
For simplicity we will assume that their are equal, $C_{A}(t)=C_{D}(t)\equiv C(t)$,
but the fluctuations on different molecules remains uncorrelated.
The absorption and fluorescence spectra, as well as non-linear optical
spectra including 2D spectra can be expressed using a double integral
of the bath correlation function, so-called line-shape function \cite{MukamelBook}
\begin{equation}
g(t)=\frac{1}{\hbar^{2}}\int\limits _{0}^{t}{\rm d}\tau\int\limits _{0}^{\tau}{\rm d}\tau^{\prime}C(\tau^{\prime}).
\end{equation}
We consider the bath correlation function, and correspondingly the
line shape function, containing two components
\begin{equation}
g(t)=g_{{\rm bath}}(t)+g_{{\rm vib}}(t),
\end{equation}
where $g_{{\rm bath}}(t)$ describes the energy gap fluctuations due
to interaction with a large harmonic bath, and $g_{{\rm vib}}(t)$
describes the contributions of underdamped oscillations due to individual
intramolecular vibrational modes. We choose to treat the case of a
single vibrational mode and we use
\begin{equation}
g_{{\rm osc}}(t)=\frac{\lambda}{\omega}(\Theta(T)(1-\cos(\omega t))+i\sin(\omega t)-i\omega t),\label{eq:gosc}
\end{equation}
where $\Theta(T)=\coth(\hbar\omega/2k_{B}T)$ (see Ref. \cite{MukamelBook}).
We will assume the bath to be characterized by Debye spectral density
with some correlation time $\tau_{{\rm bath}}$, and reorganization
energy $\lambda_{{\rm bath}}$. The line shape function $g_{{\rm bath}}(t)$
corresponding to this bath correlation function is linear at large
values of $t>\tau_{{\rm bath}}$, i.e. $g_{{\rm bath}}(t)\approx\alpha t$.
In particular for the imaginary part we have ${\rm Im}g_{{\rm bath}}(t)\approx-i\lambda t$
at long times \cite{MukamelBook}.

The energy transfer rate $K_{AD}$ from the donor to acceptor is in
principle time dependent and it can be calculated with the help of
correlation functions (line-shape functions) of the donor and acceptor
\cite{Mukamel1995a}. We assume that the part $g_{{\rm vib}}(t)$
of the energy gap fluctuation does not participate on the transfer
rate, and we assume that the rate becomes quickly time independent.
Later in this paper we study the dependence of various vibrational
features in spectroscopic signals on the magnitude of the rate $K_{AD}$
which we always assume to be constant.

\section{Non-linear Response Functions with Energy Transfer\label{sec:Non-linear-Response-Functions}}

Non-linear spectroscopic signals can be calculated conveniently using
the response function formalism \cite{MukamelBook}. In the following,
we will use the theory developed in Ref. \cite{Yang1999a} to describe
non-linear optical signal of an acceptor-donor system. In Ref. \cite{Yang1999a}
the nuclear DOF are treated via second cumulant expansion of the response
functions. For Gaussian baths, such as the bath of an infinite number
of harmonic oscillators coupled linearly to the electronic transition,
this approach leads to exact expressions for the response. Adding
a single independent vibrational mode, which is well pronounced in
the non-linear response, allows us to model the transfer of the nuclear
oscillations during energy exchange between the monomers.

\begin{figure}
\includegraphics[width=0.9\columnwidth]{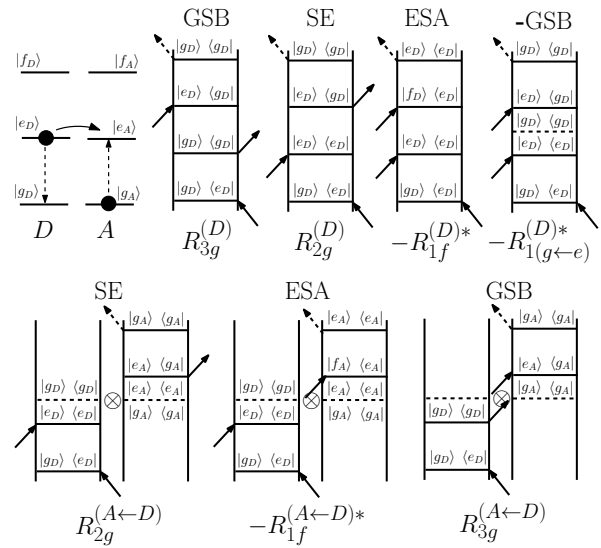}

\caption{\label{fig:Feynman-diagrams}The diagram of the levels of a weakly
coupled dimer system and the Feynman diagrams of the rephasing Liouville
pathways used in this paper.}
\end{figure}

In Ref. \cite{Yang1999a} two two-level systems connected through
a transfer rate $K_{AD}$ were studied. In our case we need to formally
include excited state absorption to higher excited state other than
the two-exciton states discussed above. However, this addition does
not change the treatment of the energy transfer, because higher excited
states do not contribute to the energy transfer dynamics observed
by third order spectroscopic methods (such as 2D coherent spectroscopy)
\cite{MukamelBook}. We will study the transfer in one direction,
from a donor molecule denoted as $D$ to an acceptor molecule denoted
as $A$. We will denote nonlinear response functions corresponding
to experimental signal by superindices $(D)$ and $(A)$, or $(A\leftarrow D)$
if the response contains energy transfer. We will show that this process
leads to a partial loss of the amplitude of the oscillations due to
vibrational coherence. 

There are four types of double-sided Feynman diagrams representing
four general types of response functions. These are usually denoted
by lower index $n=1,\dots,4$ in the literature (see e.g. \cite{MukamelBook}).
In addition, we distinguish the pathways that include the ground state
$|g\rangle$ and the excited state $|e\rangle$ only, and those involving
the higher excited state $|f\rangle$. To the former we add a lower
index $g$ while to the latter we add a lower index $f.$ Some pathways
involve relaxation from the excited state $|e\rangle$ to the ground
state $|g\rangle$ and we denote them with a lower index $(g\leftarrow e)$.
The list of rephasing pathways and their corresponding Feynman diagrams
are presented in Fig. \ref{fig:Feynman-diagrams}

In this section, without the loss of generality, we consider only
the rephasing part of the response. All calculations presented later
in Section \ref{sec:Discussion} will be based on a complete set of
pathways. We assume that originally only the donor is excited, although
in reality, the process where the role of the donor and the acceptor
are exchanged occurs simultaneously. There are seven rephasing Liouville
pathways with the following interpretations: (i) The ground state
bleach (GSB) signal from the donor is attributed to the pathway $R_{3g}^{(D)}$.
During the population time, this contribution evolves due to the bath
reorganization and the oscillation of the vibrational mode, but it
does not change due to the energy transfer process. (ii) The donor
stimulated emission (SE) which exponentially decays with the energy
transfer rate $K_{AD}$ is described by the response function $R_{2g}^{(D)}$
multiplied by the corresponding decay factor. (iii) When the donor
decays to the ground state due to energy being transferred to the
acceptor the bleach is filled with the decayed population. This process
is described by the response function $R_{1(g\leftarrow e)}^{(D)*}$
where the star denotes complex conjugation. This response function
has an overall type of an $R_{1}$ pathway, but it carries a minus
and a complex conjugation. It rises with a factor $(1-e^{-K_{AD}t_{2}})$.
The minus leads to cancellation of the GSB contribution of $R_{3g}^{(D)}$.
(iv) Together with the energy transfer there is a corresponding rise
of the GSB of the acceptor. This is described by the response function
$R_{3g}^{(A\leftarrow D)}$ and the rise is the same as in pathway
(iii). (v) and (vi) Similarly to the GSB also the SE and the ESA of
the acceptor rise. These processes are described by the response functions
$R_{2g}^{(A\leftarrow D)}$ and $R_{1f}^{(A\leftarrow D)}$, respectively.
(vii) Finally, the ESA of the donor is decaying in exactly the same
way as its SE, and it is described by the response function $R_{1f}^{(D)*}$
multiplied by the corresponding exponentially decaying factor.

\subsection{Transfer of the Excited State Vibrational Coherence}

Using the matrix elements of the Liouville space density matrix propagators
for optical coherences ${\cal U}_{eg}(t)\equiv{\cal U}_{egeg}(t)$
and the ground- and excited state density matrix elements ${\cal U}_{ee}(t)\equiv{\cal U}_{eeee}(t)$
and ${\cal U}_{gg}(t)\equiv{\cal U}_{gggg}(t)$, where we abbreviated
the number of electronic indices to two, we can express the above
discussed response functions in the following way (see e.g. \cite{Mancal2013z}):
\begin{equation}
R_{2g}^{(D)}\approx\langle{\cal U}_{e_{D}g_{D}}(t_{3}){\cal U}_{e_{D}e_{D}}(t_{2}){\cal U}_{ge}(t_{1})\rangle_{D}e^{-K_{AD}t_{2}},\label{eq:standard0}
\end{equation}
\begin{equation}
R_{3g}^{(D)}\approx\langle{\cal U}_{e_{D}g_{D}}(t_{3}){\cal U}_{g_{D}g_{D}}(t_{2}){\cal U}_{g_{D}e_{D}}(t_{1})\rangle_{D},\label{eq:standard1}
\end{equation}
\begin{equation}
R_{1f}^{(D)*}\approx-\langle{\cal U}_{f_{D}e_{D}}(t_{3}){\cal U}_{e_{D}e_{D}}(t_{2}){\cal U}_{g_{D}e_{D}}(t_{1})\rangle_{D}e^{-K_{AD}t_{2}},\label{eq:standard2}
\end{equation}
\[
R_{1(g\leftarrow e)}^{(D)*}\approx-K_{AD}\int\limits _{0}^{t_{2}}d\tau\langle{\cal U}_{e_{D}g_{D}}(t_{3}){\cal U}_{g_{D}g_{D}}(t_{2}-\tau)
\]
\begin{equation}
\times{\cal U}_{e_{D}e_{D}}(\tau){\cal U}_{g_{D}e_{D}}(t_{1})\rangle_{D}e^{-K_{AD}\tau},\label{eq:R1D_ge}
\end{equation}
\[
R_{2g}^{(A\leftarrow D)}\approx K_{AD}\int\limits _{0}^{t_{2}}d\tau\langle{\cal U}_{e_{A}g_{A}}(t_{3}){\cal U}_{e_{A}e_{A}}(t_{2}-\tau)\rangle_{A}
\]
\begin{equation}
\times\langle{\cal U}_{e_{D}e_{D}}(\tau){\cal U}_{g_{D}e_{D}}(t_{1})\rangle_{D}e^{-K_{AD}\tau},\label{eq:RAD2g}
\end{equation}
\[
R_{3g}^{(A\leftarrow D)}\approx K_{AD}\int\limits _{0}^{t_{2}}d\tau\langle{\cal U}_{e_{A}g_{A}}(t_{3}){\cal U}_{g_{A}g_{A}}(t_{2}-\tau)\rangle_{A}
\]
\begin{equation}
\times\langle{\cal U}_{g_{D}g_{D}}(\tau){\cal U}_{g_{D}e_{D}}(t_{1})\rangle_{D}e^{-K_{AD}\tau},\label{eq:RAD3g}
\end{equation}
\[
R_{1f}^{(A\leftarrow D)*}\approx-K_{AD}\int\limits _{0}^{t_{2}}d\tau\langle{\cal U}_{f_{A}e_{A}}(t_{3}){\cal U}_{e_{A}e_{A}}(t_{2}-\tau)\rangle_{A}
\]
\begin{equation}
\times\langle{\cal U}_{e_{D}e_{D}}(\tau){\cal U}_{g_{D}e_{D}}(t_{1})\rangle_{D}e^{-K_{AD}\tau}.\label{eq:RAD1f}
\end{equation}
Here, we omitted the prefactors containing transition dipole moments,
and $\langle\dots\rangle_{D}=tr_{{\rm bath}}\{\dots W_{{\rm eq}}^{(D)}\}$
correspond to the averaging over the equilibrium environmental DOF
of the donor. The sign $\langle\dots\rangle_{A}=tr_{{\rm bath}}\{\dots W_{{\rm eq}}^{(A)}\}$
represents the same averaging for the acceptor. Expressions for the
standard pathways of Eqs. (\ref{eq:standard0}), (\ref{eq:standard1})
and (\ref{eq:standard2}) are well known from the literature, see
e.g. \cite{MukamelBook}. The response of Eq. (\ref{eq:R1D_ge}) will
be treated in more detail later. Let us first address the Eqs. (\ref{eq:RAD2g})
to (\ref{eq:RAD1f}). The averaging on the donor in these equations
leads to an elimination of the dependence on the variable $\tau$
in the donor part of the response, because
\[
\langle{\cal U}_{e_{D}e_{D}}(\tau){\cal U}_{g_{D}e_{D}}(t_{1})\rangle_{D}=tr_{{\rm bath}}\{{\cal U}_{e_{D}e_{D}}(\tau){\cal U}_{g_{D}e_{D}}(t_{1})W_{eq}\}
\]
\begin{equation}
=tr_{{\rm bath}}\{{\cal U}_{g_{D}e_{D}}(t_{1})W_{eq}\}.
\end{equation}
This is because the superoperator ${\cal U}_{e_{D}e_{D}}(\tau)$ acting
on an arbitrary operator $A$ corresponds to an action of two ordinary
evolution operators
\begin{equation}
{\cal U}_{e_{D}e_{D}}(\tau)A=U_{e_{D}}(t)AU_{e_{D}}^{\dagger}(t),
\end{equation}
and the operators can be reordered in a cyclic way under the trace
operation. In addition, in $R_{3g}^{(A\leftarrow D)}$ the $\tau$
dependencecan be eliminated also on the acceptor part of the response,
because
\begin{equation}
{\cal U}_{g_{A}g_{A}}(t_{2}-\tau)W_{eq}^{(A)}=W_{eq}^{(A)}
\end{equation}
due to invariance of the equilibrium to the ground state propagation.
Correspondingly we have
\begin{equation}
R_{3g}^{(A\leftarrow D)}\approx e^{-g_{A}(t_{3})-g_{D}(t_{1})}(1-e^{-K_{AD}t_{2}}).
\end{equation}
In the bleach signal there is therefore no transfer of any dependency
of the response on $t_{2}$. Other transfer pathways can now be expressed
through the purely acceptor pathways which have a standard form, with
the first time argument (from the right) equal to zero
\begin{equation}
R_{2g}^{(A\leftarrow D)}\approx e^{-g_{D}(t_{1})}K_{AD}\int\limits _{0}^{t_{2}}d\tau R_{2g}^{(A)}(t_{3},t_{2}-\tau,0)e^{-K_{AD}\tau},\label{eq:R2gAD_via_R2gA}
\end{equation}
\begin{equation}
R_{1f}^{(A\leftarrow D)*}\approx-e^{-g_{D}(t_{1})}K_{AD}\int\limits _{0}^{t_{2}}d\tau R_{1f}^{(A)*}(t_{3},t_{2}-\tau,0)e^{-K_{AD}\tau}.\label{eq:R1fAD_viaR1fA}
\end{equation}
For the stimulated emission this leads to
\[
R_{2g}^{(A\leftarrow D)}\approx e^{-g_{D}(t_{1})-g_{A}^{*}(t_{3})}
\]
 
\begin{equation}
\times K_{AD}\int\limits _{0}^{t_{2}}d\tau\ e^{2i{\rm Im}[g_{A}(t_{2}-\tau)-g_{A}(t_{2}+t_{3}-\tau)]-K_{AD}\tau},
\end{equation}
which can be relatively easily evaluated. The ESA contribution depends
on the particular assumptions we make about the higher excited states,
and cannot therefore be evaluated without introducing further assumptions.
We can consider this contribution to be similar in oscillatory features
to the SE, because the source of the oscillation is the same excited
state of the acceptor.

\subsection{Refilling of the Bleaching Signal}

The ground state bleach contribution to the rephasing signal, which
we denote by $R_{3g}^{(D)}$, appears stationary, and, at the first
sight, unaffected by the energy transfer process. However, the transfer
process is accompanied by deexcitation of the donor domain, and correspondingly,
there is a signal countering the one of the $R_{3g}^{(D)}$ pathway.
Above we denoted this signal by $R_{1(g\leftarrow e)}^{(D)*}$ above.
The diagram of this pathway has a form of the $R_{1}$ diagram mirror
imaged, and it is used with the minus sign. According to Yang and
Fleming, Ref. \cite{Yang1999a}, this pathway reads
\[
R_{1(g\leftarrow e)}^{(D)*}(t_{3},t_{2},t_{1})=R_{3g}^{(D)}(t_{3},t_{2},t_{1})e^{-2i{\rm Im}(g_{D}(t_{2}+t_{3})-g_{D}(t_{2}))}
\]
\begin{equation}
\times K_{AD}\int\limits _{0}^{t_{2}}{\rm d}t^{\prime}e^{-K_{AD}t^{\prime}}e^{2i{\rm Im}(g_{D}(t_{2}+t_{3}-t^{\prime})-g_{D}(t_{2}-t^{\prime}))}.\label{eq:R1Dtr_def}
\end{equation}
The signal canceling the GSB contains the GSB response function $R_{3g}^{(D)}$.
Later in this paper, we will evaluate all the response functions numerically.
Let us however attempt a slightly more involved analysis of Eq. (\ref{eq:R1Dtr_def})
assuming its long $t_{2}$ approximation. Because the $g_{{\rm bath}}$
and the $g_{{\rm osc}}$ components of the total donor line-shape
function are independent, one can factorize the response function
into the bath- and the vibrational parts. In addition, the bath line
shape function $g_{{\rm bath}}(t)$ is linear at its arguments larger
than $\tau_{{\rm bath}}$ and correspondingly the $t^{\prime}$ dependence
on the bath part of the line-shape function vanishes. With these assumptions,
we can write for $t_{2}>\tau_{{\rm bath}}$
\[
R_{1(g\leftarrow e)}^{(D)*}(t_{3},t_{2},t_{1})\approx R_{3g}^{(D)}(t_{3},t_{2},t_{1})e^{-2i{\rm Im}(g_{{\rm osc}}(t_{2}+t_{3})-g_{{\rm osc}}(t_{2}))}
\]
 
\begin{equation}
\times K_{AD}\int\limits _{0}^{t_{2}}{\rm d}t^{\prime}e^{-K_{AD}t^{\prime}}e^{2i{\rm Im}(g_{{\rm osc}}(t_{2}+t_{3}-t^{\prime})-g_{{\rm osc}}(t_{2}-t^{\prime}))},\label{eq:R1Dtr_osc}
\end{equation}
i.e. the influence of the bath evolution is completely hidden in the
$R_{3g}^{(D)}(t_{3},t_{2},t_{1})$ function. The integral in Eq. (\ref{eq:R1Dtr_osc})
we see only contributions originating from the intramolecular vibrations. 

Integrating Eq. (\ref{eq:R1Dtr_osc}) by parts we obtain\begin{widetext}
\[
R_{1(g\leftarrow e)}^{(D)*}(t_{3},t_{2},t_{1})\approx R_{3g}^{(D)}(t_{3},t_{2},t_{1})(1-e^{-K_{AD}t_{2}}e^{-2i{\rm Im}(g_{{\rm osc}}(t_{2}+t_{3})-g_{{\rm osc}}(t_{2})-g_{{\rm osc}}(t_{3}))})
\]
\[
-2i\frac{\lambda}{\omega}R_{3g}^{(D)}(t_{3},t_{2},t_{1})e^{-2i{\rm Im}(g_{{\rm osc}}(t_{2}+t_{3})-g_{{\rm osc}}(t_{2}))}
\]
\begin{equation}
\times\int\limits _{0}^{t_{2}}{\rm d}t^{\prime}\omega[{\rm cos}\omega(t_{2}+t_{3}-t^{\prime})-{\rm cos}\omega(t_{2}-t^{\prime})]e^{-K_{AD}t^{\prime}}e^{2i{\rm Im}(g_{{\rm osc}}(t_{2}+t_{3}-t^{\prime})-g_{{\rm osc}}(t_{2}-t^{\prime}))}.\label{eq:difference}
\end{equation}
Under the integral we used explicitly the form of the $g_{osc}$,
Eq. (\ref{eq:gosc}). At long times $t_{2}$, when $e^{-K_{AD}t_{2}}\approx0$,
the donor contribution to the overall signal consists only of the
integral term
\[
S^{(D)}(t_{3},t_{2}>K_{AD}^{-1},t_{1})=R_{3g}^{(D)}(t_{3},t_{2},t_{1})-R_{1(g\leftarrow e)}^{(D)*}(t_{3},t_{2},t_{1})
\]
\[
\approx-2i\frac{\lambda}{\omega}R_{3g}^{(D)}(t_{3},t_{2},t_{1})e^{-2i{\rm Im}(g_{{\rm osc}}(t_{2}+t_{3})-g_{{\rm osc}}(t_{2}))}
\]
\begin{equation}
\times\int\limits _{0}^{t_{2}}{\rm d}t^{\prime}\omega[{\rm cos}\omega(t_{2}+t_{3}-t^{\prime})-{\rm cos}\omega(t_{2}-t^{\prime})]e^{-K_{AD}t^{\prime}}e^{2i{\rm Im}(g_{{\rm osc}}(t_{2}+t_{3}-t^{\prime})-g_{{\rm osc}}(t_{2}-t^{\prime}))}.\label{eq:}
\end{equation}
\end{widetext}We can see that the difference between the original
GSB signal and the signal coming from the filling of the GSB is proportional
to the Huang-Rhys factor with a factor in a form of an integral over
oscillating function. The whole factor will obviously be an oscillating
function of $t_{2}$. Its numerical analysis for the range of Huang-Rhys
factors between zero and one shows that its has a leading imaginary
contribution. Correspondingly, the GSB signal on a donor does not
vanish after the excitation leaves the donor molecule. The remaining
signal is similar to the ground state contribution $R_{3g}^{(D)}$
multiplied by $i$ and modulated by oscillating real function. The
numerical results presented in Section \ref{sec:Discussion} confirm
this conclusion.

\begin{figure}
\includegraphics[width=8.5cm]{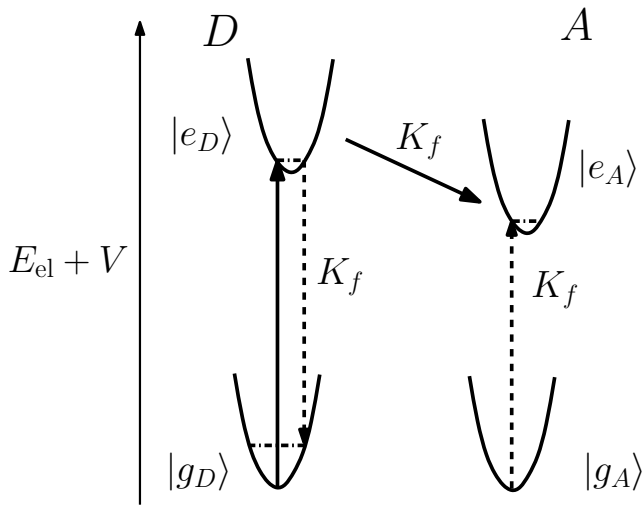}

\caption{\label{fig:Potential-energy-surfaces}Potential energy surfaces of
the dimer model with F\"{o}rster regime of energy transfer. The donor
molecule $D$ is initially excited and the excitation of is transferred
with the transfer constant $K_{F}$ to the acceptor molecule $A$.
This process corresponds to a deexcitation of the donor and excitation
of the acceptor with the same rate $K_{F}$. The excited state wavepacket
is projected to a non-equilibrium position of the ground state during
the energy transfer to the acceptor.}
\end{figure}

\subsection{Origin of the Bleach Signal}

Interestingly, after an ideal impulsive excitation of a molecular
system, no dynamic GBS (i.e. dependent on $t_{2}$) would arise. This
can be seen directly from the form of the response function while
setting $t_{1}=0$, or from an intuitive picture summarized in Fig.
\ref{fig:Potential-energy-surfaces}. In terms of the response function,
the beach corresponds to an excitation and then deexcitation of one
of the sides of the Feynman diagram (see pathway $R_{3g}^{(D)}$ of
Fig. \ref{fig:Feynman-diagrams}). The beating which we observe in
$t_{2}$ is then the beating between the static unexcited ground state
wavepacket (or more precisely an equilibrium mixed state) and a perturbed
wavepacket which was excited, evolved for a short time $t_{1}$ on
the excited state PES, and was then transferred back to the ground
state into a non-equilibrium position. If the delay $t_{1}=0$, the
wavepacket does not have time to evolve on the excited state PES and
is returned back to its original equilibrium state. Hence no dynamical
signal arises. 

During the excitation transfer, the GSB signal is compensated by an
excited state population returning to the ground state. In the non-linear
response corresponding to the excited states there are in principle
two wavepackets (in the ket and the bra of the Feynman diagram) which
were created in the excited state at two times occurring with a delay
$t_{1}$. These wavepackets are projected to the ground state of the
donor during the excitation transfer, and they are unlikely to have
a phase as to cancel the oscillations that already occur in the ground
state. It is therefore not surprising that the GSB signal will remain
after the excitation is transferred. In the next section we will study
the amplitude of these remaining oscillations and its dependence on
the value of the energy transfer rate.

\section{Discussion\label{sec:Discussion}}

\begin{figure}
\includegraphics[width=8.5cm]{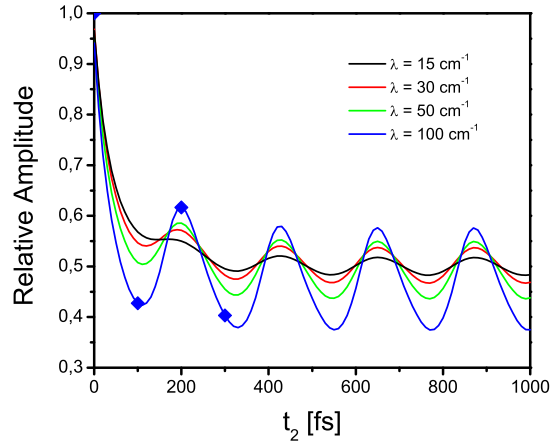}

\caption{\label{fig:Oscillations-underdamped}Oscillations due to underdamped
vibrations of nuclear mode with frequency $\omega=150$ cm$^{-1}$
for various reorganization energies. The full blue diamonds denote
the positions for which real part of the 2D spectrum is plotted in
Fig. \ref{fig:2D-1}.}
\end{figure}

\begin{figure}
\includegraphics[width=8.5cm]{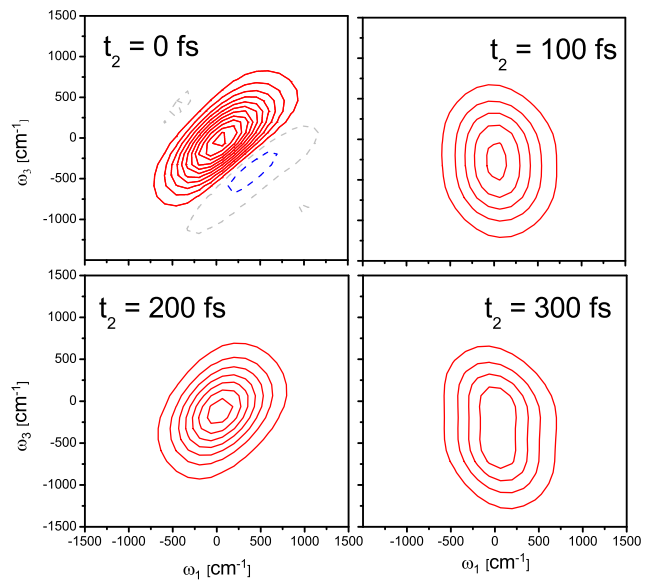}

\caption{\label{fig:2D-1}The time evolution of the real part of the 2D coherent
spectrum of a donor molecule in the absence of energy transfer to
an acceptor. The snapshots are taken at $t_{2}=0,100,200$ and $300$
fs. The reorganization energy $\lambda$ of the vibrational mode is
$100$ cm$^{-1}$. The figures correspond to the positions on a blue
curve in Fig. \ref{fig:Oscillations-underdamped} which are marked
by full blue diamonds. The figures are normalized to maximum of the
2D spectrum at $t_{2}=0$ fs, and there are $12$ contours between
zero and the maximum. }
\end{figure}

In this section, we will present results of numerical calculations
of 2D coherent Fourier transformed spectra. The definition of the
2D spectrum as well as the description of the experimental technique
can be found e.g. in \cite{Jonas2003a,Brixner2004a}. Throughout this
section we use one parameters set for the bath correlation function,
namely, $\lambda_{{\rm bath}}=200$ cm$^{-1}$, $\tau_{{\rm bath}}=100$
fs. The temperature is assume to be $T=300$ K. These parameters ensure
a realistically broad 2D spectrum on which a typical frequency of
an intramolecular vibrational mode, $\omega=150$ cm$^{-1}$ and an
oscillator reorganization frequency $\lambda<\omega$ lead to characteristic
line shape modulation (see e.g. \cite{Nemeth2008a}). It is customary
to present the real part of the spectrum, which corresponds to the
absorption--absorption/stimulated emission plots \cite{Jonas2003a,Mancal2013z}.
Similarly to the situation in chlorosome, the presence of the vibrational
mode does not lead here to any discernible crosspeak, but the spectral
amplitude at its center, and the overall line shape are modulated.
Fig. \ref{fig:Oscillations-underdamped} shows the characteristic
oscillations of the amplitude of the 2D spectrum at the resonance,
for the donor molecule with no excitation energy transfer ($K_{AD}=0$).
Fig. \ref{fig:2D-1} presents the corresponding 2D spectra at some
selected points (denoted by diamonds in Fig. \ref{fig:Oscillations-underdamped}).
All 2D spectra are represented with respect to the resonant optical
transition frequency, and they are therefore centered around the $(0,0)$
point. On both figures we can notice the initial drop of the amplitude
and broadening, which occurs with the time-scale of the bath correlation
time $\tau_{{\rm bath}}$. For delay times longer than $\tau_{{\rm bath}}$
the 2D spectrum shows a characteristic pattern of line shape oscillation
which corresponds to an oscillation of the amplitude. The oscillations
are more pronounced with increasing reorganization energy of the oscillator,
i.e. with increasing Huang-Rhys factor. 

\begin{figure}
\includegraphics[width=8.5cm]{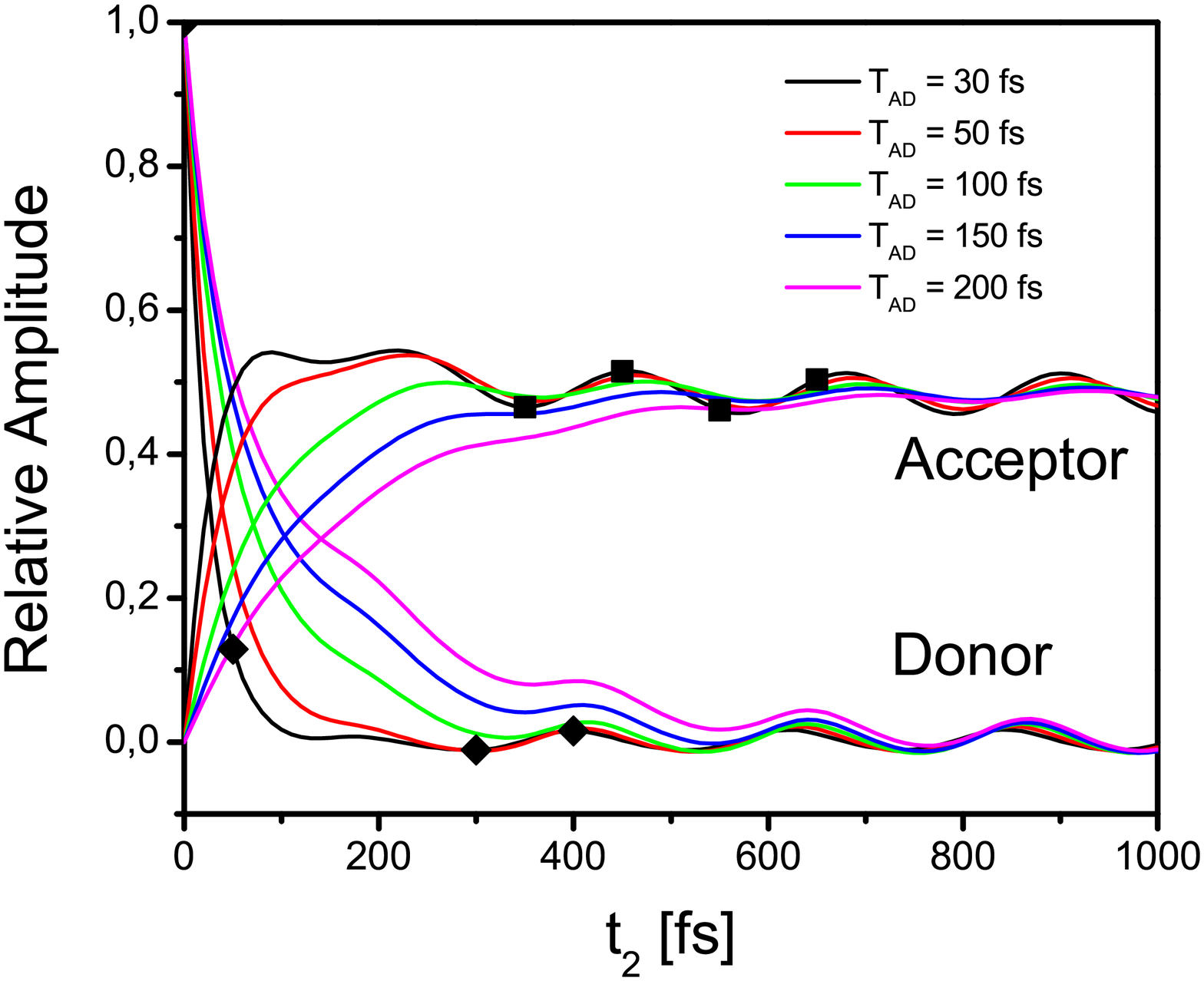}

\caption{\label{fig:Oscillations-DA}Oscillations of the 2D spectrum at the
optical resonance. Amplitude of the spectrum at $\omega_{1}=\omega_{3}$
is plotted for the donor molecule and for the acceptor molecule for
different values of the transfer time $T_{AD}=1/K_{AD}$ and the reorganization
energy $\lambda=100$ cm$^{-1}$. The vibrational mode has a frequency
$\omega=150$ cm$^{-1}$. The full black diamonds indicate the positions
for which the real part of the 2D spectrum is plotted in Fig. \ref{fig:Bleaching-signal},
and full back squares indicate the positions for which the real part
of the 2D spectrum is plotted in Fig. \ref{fig:Transferred-coherences}.}
\end{figure}

\begin{figure}
\includegraphics[width=8.5cm]{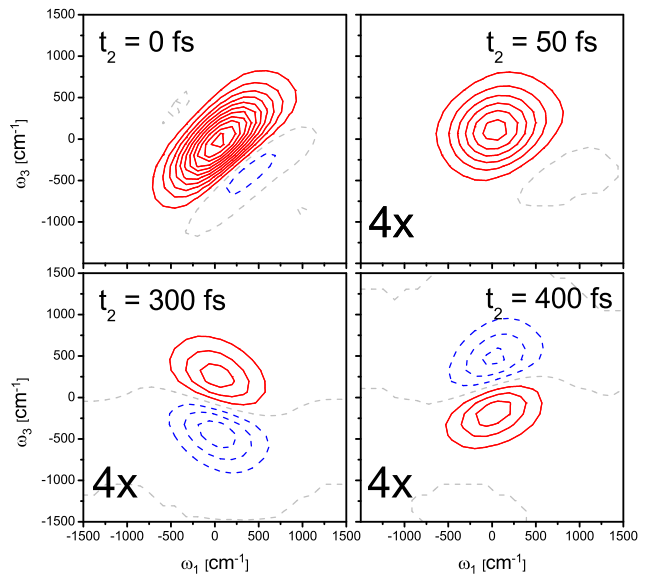}

\caption{\label{fig:Bleaching-signal}Decay of the signal on the donor. The
short time behavior ($t_{2}=0$ and $50$ fs) corresponds to the bath
dephasing and the loss of population. When the population is completely
lost and the system has relaxed back to the groundstate electronically,
the mismatch between the returning excited state nuclear wavepacket
and the groudstate wavepacket yields a dispersion patern known from
the imaginary part of the 2D spectra. The dispersion patern changes
sign with the period of the oscillation. The snapshots correspond
to the positions on the black curve in Fig. \ref{fig:Oscillations-DA}
which are marked by black diamonds. The figures are normalized to
maximum of the 2D spectrum at $t_{2}=0$ fs, and there are $12$ contours
between zero and the maximum. At times $t_{2}=50,\ 300,$ and $400$
fs, we use four times as many contours as in the $t_{2}=0$ fs spectrum
to enhance the small amplitude of the spectrum.}
\end{figure}

\begin{figure}
\includegraphics[width=8.5cm]{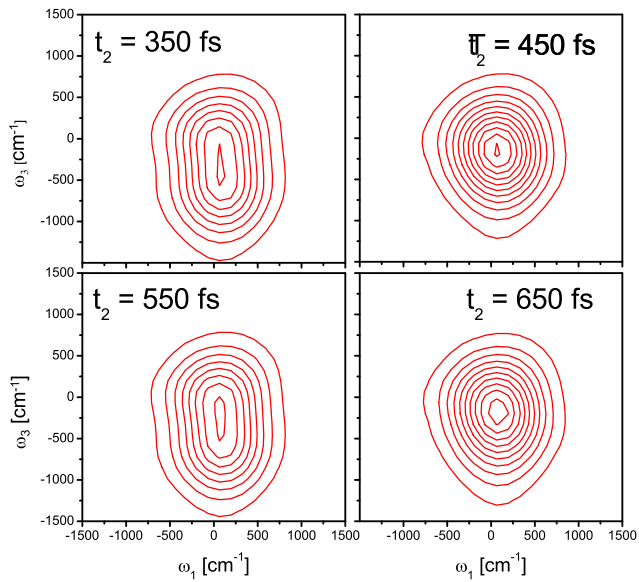}

\caption{\label{fig:Transferred-coherences}Time evolution of coherences transferred
to the acceptor molecule. After the rise of the 2D signal is completed
(see Fig. \ref{fig:Oscillations-DA}), the spectrum keeps evolving
due the nuclear motion in the excited state of the acceptor molecule.
The figures are normalized to the maximum of 2D spectrum at $t_{2}=450$
fs (largest amplitude of the four spectra), and there are $12$ contours
between zero and its maximum.}
\end{figure}

\begin{figure}
\includegraphics[width=8.5cm]{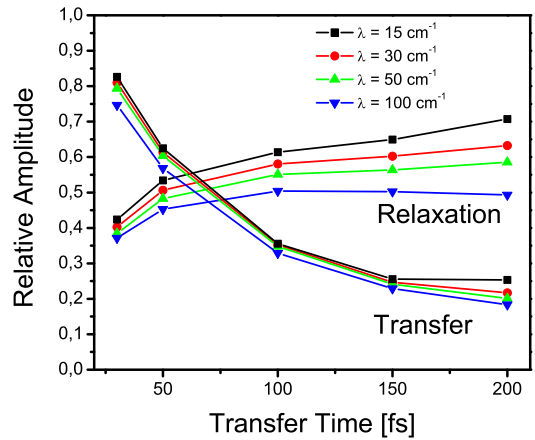}

\caption{\label{fig:Effciency}Efficiency of the coherence transfer. Curves
doted as ``Transfer'': Ratio of the amplitude transferred to the
acceptor to the original amplitude of the oscillations on the donor.
Curves denoted as ``Relaxation'': Ratio of the amplitude of the
ground state bleaching oscillations on the donor to the original amplitude
of the oscillations on the donor. All parameters are same as in Fig.
\ref{fig:Oscillations-DA}.}
\end{figure}

\begin{figure}
\includegraphics[width=8.5cm]{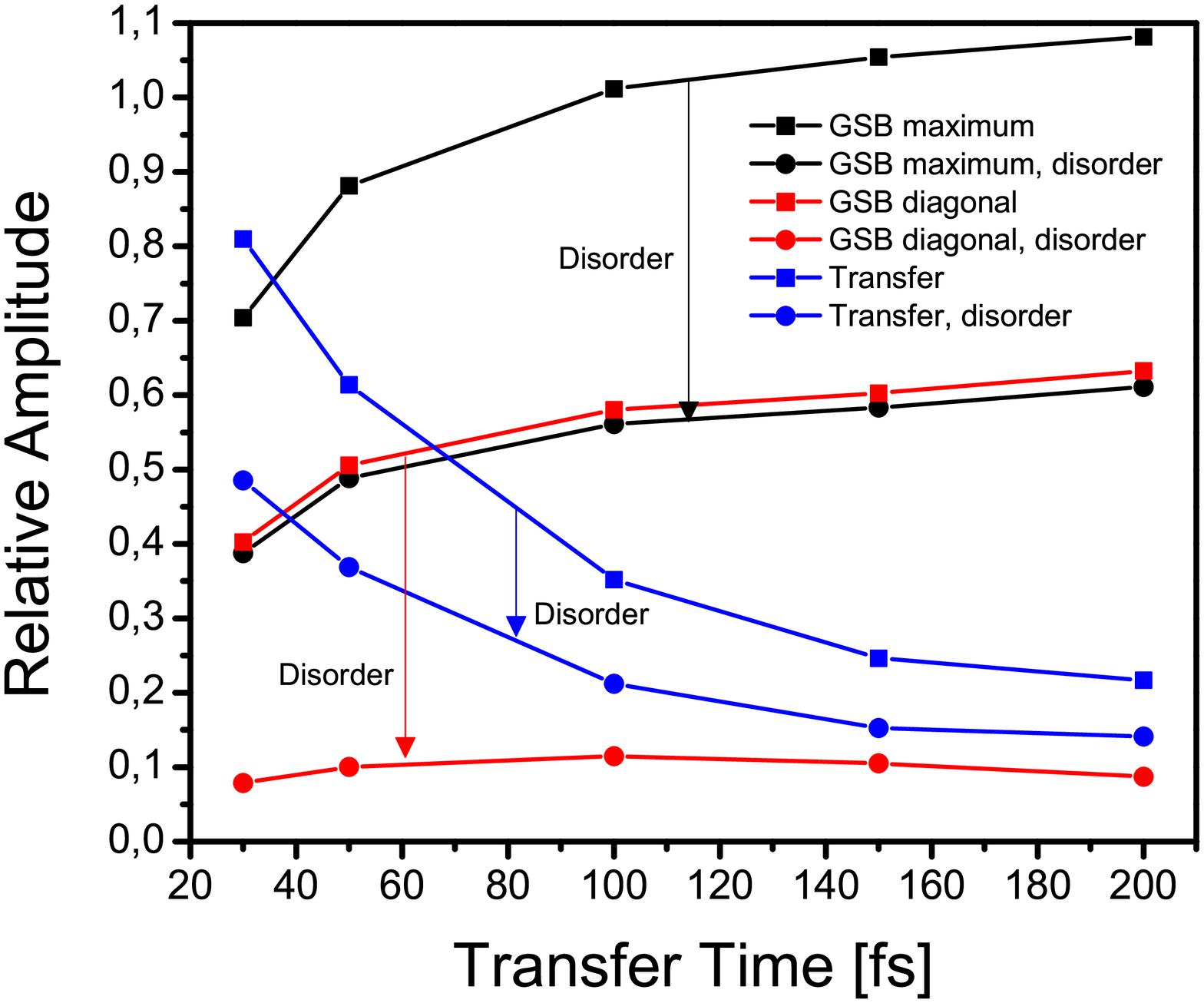}

\caption{\label{fig:Influence-of-disorder}Influence of electronic disorder
on the relative amplitude of the oscillations with frequency $\omega=150$
cm$^{-1}$ and reorganization energy $\lambda=30$ cm$^{-1}$. The
disorder corresponding to a distribution of acceptor and donor transition
frequencies with full width at half maximum (FWHM) equal to $1000$
cm$^{-1}$ was applied. The relative amplitude of the transferred
coherences (in blue) is diminished only slightly. At the diagonal
of the 2D spectru, the groud state bleaching oscillations are reduced
by more than a factor of $5$ (in red). However, when the amplitude
is measured where the bleaching spectrum has its maximum (see lower
panels of Fig. \ref{fig:Bleaching-signal}), the relative amplitude
is decreased only by factor of $2$, and the beating on the relaxed
donor remains larger than the one on the acceptor. }
\end{figure}

When the energy transfer from the donor to the acceptor is allowed,
the total donor signal is quickly disappearing as the stimulated emission
pathway decays exponentially and the ground-state bleach is compensated
by the signal from the population arriving to the ground state from
the excited state. Fig. \ref{fig:Oscillations-DA} shows, however,
that the donor signal remains oscillating around zero. The signal
on the acceptor, on the other hand, rises to an amplitude corresponding
to the one exhibited by the donor without energy transfer (cf. Fig.
\ref{fig:Oscillations-underdamped}). One can notice by eye that the
amplitude of the oscillations decreases with the increasing transfer
time (decreasing transfer rate). One can also notice that this is
not the case for the ground state bleach. Characteristic 2D spectra
of the donor contribution are presented in Fig. \ref{fig:Bleaching-signal}.
Originally almost completely positive signal quickly disappears on
the timescale of the energy transfer, and what remains is a signal
with a characteristic shape of the imaginary part of the 2D spectrum
(see e.g. \cite{Nemeth2008a}), but with an alternating sign. The
rising acceptor signal, Fig. \ref{fig:Transferred-coherences}, shows
only the characteristic line shape modulations accompanied with the
change of the maximum amplitude of the spectrum. 

The relative amplitude of the coherent oscillations transferred to
the acceptor or left on the donor with respect to the amplitude expected
on the non-transferring donor is plotted in Fig. \ref{fig:Effciency}.
We compare relative amplitudes for four different values of the oscillator
reorganization energy. We find that the transfer of oscillations to
the acceptor for all reorganization energies follows roughly the same
exponential decay with the transfer time. This behavior is expected,
because a slow feeding rate leads to more destructive interference
between various contributions of the transferred nuclear wavepacket.
For electronic coherence this effect was recently discussed in Ref.
\cite{Chenu2013b}. Similar behavior is observed when electronic coherence
is induced a multilevel molecular system by pulsed light with increasing
duration as shown by Jiang and Brumer in Ref. \cite{Jiang1991a}.
This trend is however not followed by the GSB signal remaining on
the donor. Fig. \ref{fig:Effciency} demonstrates that the relative
amplitude of the oscillations on the resonance does not decay with
increasing transfer time. It even slightly increases at short times,
and it stays flat for times up to 200 fs. For longer transfer times,
the wavepacket stays for a longer time in the excited state. Because
the period of the oscillator is larger than the transfer time, it
apparently acquires a larger amplitude when it is projected on the
ground state.

Real molecular systems are studied by the non-linear spectroscopy
in form of macroscopic disordered ensembles. The chlorosome which
motivates our study of the transfer of vibrational coherence is also
a strongly disordered system. It is therefore worth studying the effect
of the transition energy disorder on the oscillatory patter observed
in 2D spectra. In general, a change in the transition energy results
in a displacement of the 2D line shape along the diagonal line of
the spectrum. The disorder thus corresponds to an additional elongation
of the line shape along the diagonal line. For a line shape which
is positive everywhere, this does not mean any significant change
in the total amplitude of the spectrum (it should get slightly diminished
due to the broadening). For a line shape corresponding to the times
$t_{2}=300$ fs and $t_{2}=400$ fs in Fig. \ref{fig:Bleaching-signal},
however, the displacement may lead to additional canceling of the
signal from differently displaced line shapes, because the line shapes
contain both positive and negative regions. In Fig. \ref{fig:Influence-of-disorder}
we therefore study the influence of the disorder on the amplitude
of the transferred oscillations and on the amplitude of the remaining
GSB signal. In our calculations we assume a large Gaussian disorder
simulated by a normal distributions of the transition energies with
a full width at half maximum of $1000$ cm$^{-1}$. In all studied
cases the disorder leads to a decrease of the relative amplitude of
the oscillations. For the oscillation transfer this decrease is less
than 50 \%. Similarly for the points of the 2D spectrum where the
GSB has its maximum at $t_{2}=300$ fs, the decay of the amplitude
of the oscillations is less than 50 \%. On the diagonal, i.e. near
the nodal line of the 2D spectrum, the amplitude is decreased almost
five times. Nevertheless, the donor signal remains oscillating despite
disorder, and the expected amplitude of the oscillations in comparable
to or larger than the amplitude of the transferred oscillations.

From the point of view of large aggregates the results concerning
the survival of the bleach signal are the most interesting. In system
where each donor has a number of acceptors, the excitation which started
on a given donor will be very rarely detected returning to the donor
(after it has passed through some other molecules). Even the process
in which the excitation passed through its original donor and was
then detected on some of the acceptors is much less frequent than
the situation in which the excitation leaves the donor and never returns
or passes through it. In such a case, the bleaching signal remains
oscillating until the vibrational energy relaxes due to processes
independent of the energy transfer process. In the same situation,
continuing transfer of the excitation to subsequent acceptors will
lead to a complete decay of the nuclear oscillations originating in
the excited state, i.e. oscillations in the SE and ESA signals.

\subsection{Outlook}

The discussion in this paper is only a start of a more extensive research
program, which has to incorporate several important effects which
were neglected here. First, one has to take into account the excitonic
character of the states if one wants to draw some conclusions for
the possible vibrational coherence transfer in photosynthetic systems
in general. Also, various transition dipole moment borrowing effects
including formation of so-called vibronic (vibrational-excitonic)
excitons have to be considered in case where resonance between the
energy gap in the heterodimer and the vibrational frequency occurs.
These two different effects amount to a study of different model situation.
In the present model, however, we have also included several approximations
which can be closely investigated. For instance, we assumed constant
energy transfer rates, i.e. we assumed certain coarse graining of
our problem in time. For fast vibrations, the relaxation rates might
still be time dependent during several first periods of the vibrational
motion after excitation. The relaxation rates might also be dependent
on the vibrational motion itself during this initial time interval.
Such effects may change the situation both quantitatively and qualitatively
in some cases, and will be studied elsewhere.

\section{Conclusions\label{sec:Conclusions}}

We have investigated the stimulated emission and ground state bleach
signals of a molecular dimeric system with pronounced vibrational
modes. We have concentrated on the transfer of the vibrational coherence
between an excited donor molecule and its neighboring acceptor. Using
response function formalism adapted for the case of a resonant energy
transfer with constant rates, we find that the nuclear oscillations
can be transferred to a neighboring molecules. Their amplitude, however,
decays with decreasing transfer rates. On the acceptor, the oscillations
are solely due to the nuclear wavepacket in the electronically excited
state. Interestingly, the amplitude of oscillations which prevail
on the donor, after it was deexcited due to energy transfer, do not
decrease with the decreasing energy transfer rate. Amplitude of both
types of oscillations are decreasing in the presence of electronic
disorder. In systems where excitation travels away from the original
donor, the dominating and surviving contribution after many steps
will be the one originating from the electronic ground state of the
donor molecule. 
\begin{acknowledgments}
This work was funded by the Czech Science Foundation (GACR) grant
no. 205/10/0989. 
\end{acknowledgments}
\bibliographystyle{prsty}

\end{document}